\documentclass[aps,prl,reprint,amsmath,amssymb,10pt]{revtex4-1}

\usepackage{amssymb,amsfonts,amsmath,graphicx}
\usepackage{color}

\usepackage{fullpage}
\usepackage{amssymb}
\usepackage{amsmath}
\usepackage{graphicx}
\usepackage{subfigure}
\usepackage{mdframed}
\usepackage{mathtools}
\usepackage{mathrsfs}
\def\be{\begin{equation}}
\def\ee{\end{equation}}
\def\bea{\begin{eqnarray}}
\def\eea{\end{eqnarray}}

\def\p{\partial} 
\def\nn{\nonumber}
\def\f{\frac}

\def\l{\left(}
\def\r{\right)}

\newcommand{\me}{\mathrm{e}}

\newcommand{\Lagr}{\mathcal{L}}

\DeclareMathOperator*{\argmin}{arg\,min} 

\usepackage[
margin=0.9in,
]{geometry}

\begin{document}

\title{Efficiency of a Stochastic Search with Punctual and Costly Restarts }

 \author{Kabir Husain}
 \affiliation{%
 The Simons Centre for the Study of Living Machines, National Centre for Biological Sciences (TIFR), Bellary Road, Bangalore 560 065, India
 }%
 \author{Sandeep Krishna}
 \affiliation{%
 The Simons Centre for the Study of Living Machines, National Centre for Biological Sciences (TIFR), Bellary Road, Bangalore 560 065, India
 }%

\date{\today}

\begin{abstract}
The mean completion time of a stochastic process may be rendered finite and minimised by a judiciously chosen restart protocol, which may either be stochastic or deterministic. Here we study analytically an arbitrary stochastic search subject to an arbitrary restart protocol, each characterised by a distribution of waiting times. By a direct enumeration of paths we construct the joint distribution of completion time and restart number, in a form amenable to analytical evaluation or quadrature; thereby we optimise the search over both time and potentially costly restart events. Analysing the effect of a punctual, i.e. almost deterministic, restart, we demonstrate that the optimal completion time always increases proportionately with the variance of the restart distribution; the constant of proportionality depends only on the search process. We go on to establish simple bounds on the optimal restart time. Our results are relevant to the analysis and rational design of efficient and optimal restart protocols.
\end{abstract}

\maketitle


Stochastic searches, in which a target of interest is located by a random process, are ubiquitous in both the natural \cite{RaphaelReview} and the computer sciences \cite{lubylasvegas}. They may be found repeatedly in biology across a range of length scales, from the reaction kinetics of proteins in complex environments to the behavioural patterns of foragers \cite{RaphaelReview,GrillResets,CircularDNA}. These situations admit a natural interpretation as first passage time (FPT) or completion time problems \cite{VanKampen}, whose efficiency and speed is biologically or algorithmically desirable. Among several optimisation strategies, recent years has seen an interest in analysing the consequences of a `restart' mechanism on the search process, in which the system is subject to a stochastic (or deterministic) restart while searching for the target \cite{EvansMajumdarPRL,EvansMajumdarOptimal,EvansMajumdarOptimalNonEq,InterfacesResetting,ResettingToTheMax,PalPotentialDiffusion,SanjibDeviations,PowerLawResets,PalTimeDepResets,EuleMetzger,SeifertThermo,Redner,MMRS_Reuveni,ReuveniUniversal,PalReuveni}. Remarkably, a diverging mean completion time may be rendered finite by the introduction of a restart protocol \cite{EvansMajumdarPRL,EvansMajumdarOptimal}.

\indent This has led to the question of how to optimise a particular search by a judiciously chosen restart process. While the bulk of existing work has considered Poisson \cite{EvansMajumdarPRL,EvansMajumdarOptimal,EvansMajumdarOptimalNonEq,InterfacesResetting,ResettingToTheMax,SanjibDeviations,PalPotentialDiffusion,ReuveniUniversal}, power-law distributed \cite{PowerLawResets} or deterministic \cite{Redner,PalTimeDepResets} restart protocols, there exists a large, unexplored space of possible restart mechanisms \cite{PalTimeDepResets,PalReuveni,EuleMetzger}; their study by direct enumeration is daunting. Recent work has establised that deterministic restarting \textit{globally} minimises the mean completion time \cite{PalTimeDepResets,PalReuveni}, but physical constraints on the restart process (e.g. unavoidable stochastic fluctuations in biological systems) might make implementation of this optimal protocol unviable. Therefore, it is important to understand how introducing stochasticity into the restart process translates into changes in the mean completion time. Further, while minimising the search time may sometimes be desirable, others costs - for example, constraints on the total number of required restart events - would demand a different protocol, and this too has remained unexplored.

\indent Here, we develop a simple yet general formalism to address these questions for a wide class of search and restart problems; we do so by solving exactly for the Laplace-transformed joint distribution of completion times and restart number. Using this result we go on to study: (i) the effect of fluctuations in the restart process on the optimal completion time, (ii) the efficiency of restart protocols when each restart event invokes a cost, and (iii) simple bounds on the globally optimal restarting time.

\begin{figure}[b]
\begin{centering}
\includegraphics[width=8cm]{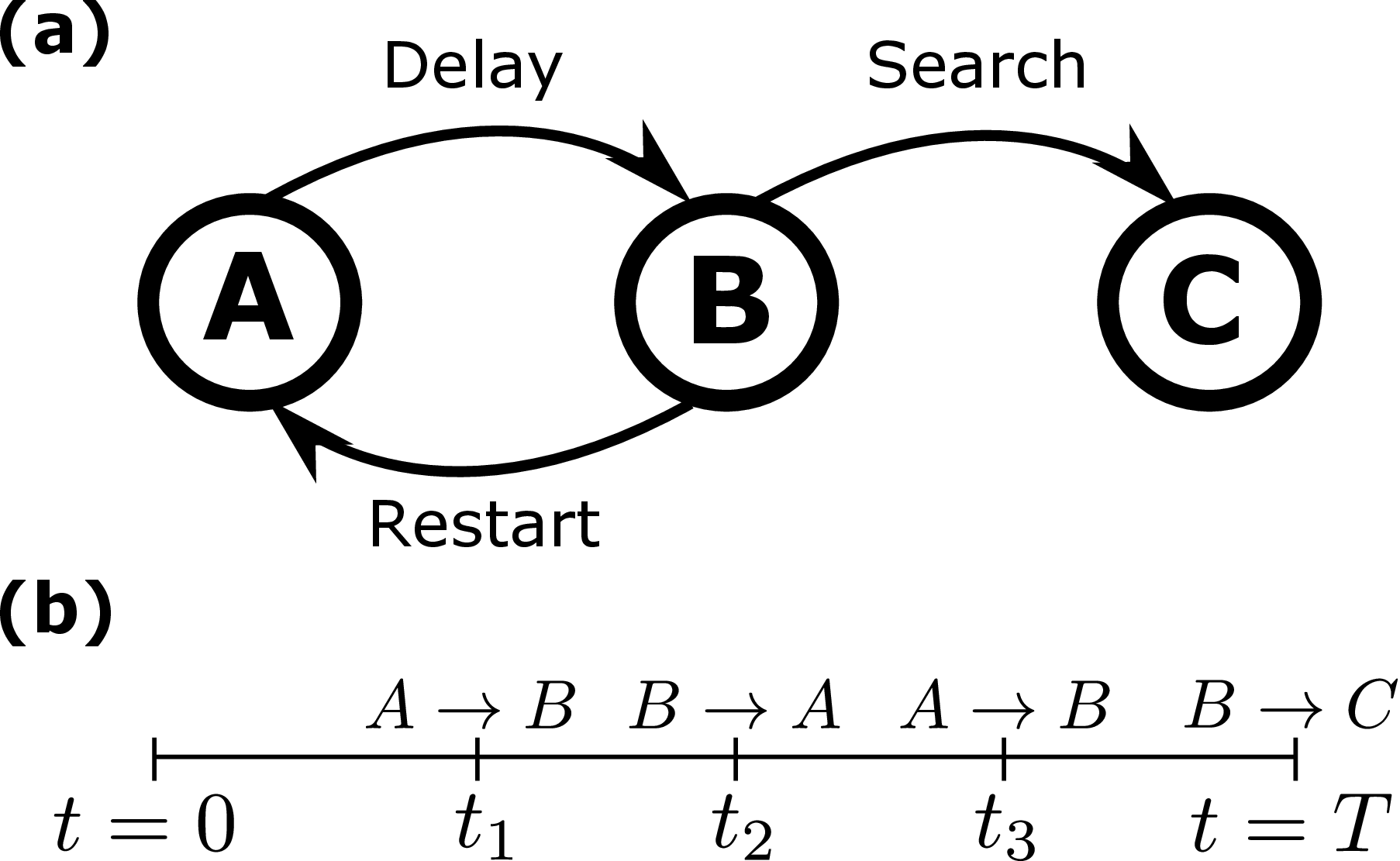}
\caption{\label{fig:schematic} Schematic of a general search process with restarts. \textbf{(a)} A `Michaelis-Menten' representation of a stochastic search with restarts \cite{MMRS_Reuveni}. \textbf{(b)} A particular path of the system from state $A$ at $t=0$ to state $C$ at $t=T$, with a single ($m=1$) restart transition ($B\to A$).}
\end{centering}
\end{figure}

\indent We obtain our results by considering the three state system depicted in Figure \ref{fig:schematic}(a). This representation, which encapsulates a large class of restart problems (but not all, for example \cite{ResettingToTheMax}), was first analysed in the context of restart processes in \cite{MMRS_Reuveni,ReuveniUniversal,PalReuveni} by way of a renewal theory framework. Here, we undertake a complementary analysis to solve for the joint distribution of completion times and restart number.

\indent At $t=0$ the system begins in state $A$ and at $t=T$ first transitions into state $C$. We identify the transition $B \to C$ with that of the search ($s$) process, $B\to A$ with the restart ($r$) and $A \to B$ as a stochastic delay process ($d$) \cite{ReuveniUniversal}. Each of these transitions is associated with its own waiting time distribution, $P_{x}(t)$ ($x$ being $s$, $r$ or $d$), the form of which characterises the process under study. 

\indent It is instructive to give a concrete example: in the context of the widely studied case of 1D diffusion with Poisson restarts \cite{EvansMajumdarPRL,EvansMajumdarOptimal}, $P_s(t)$ would be the Levi-Smirnov distribution: $\f{\me^{-1/t}}{\sqrt{\pi} t^{3/2}}$, $P_d(t)$ would be instaneous ($=\delta(t)$) and $P_r(t)$ would be an exponential distribution with constant rate $k$: $k\,\exp (-kt)$. While we shall use this specific example later to illustrate our results, we will first deal with the general case, assuming only that the $P_x(t)$ are normalised (and therefore normalisable) in the usual way, but are otherwise arbitrary.

\indent We compute the distribution of completion times $P(T)$ by an explicit summation over paths (for a recent review of similar approaches, see \cite{FreySummation}). Consider first a particular path the system takes from $A$ to $C$, wherein it undergoes the restart transition $B \to A$ $m$ times, $m \in \{ 0,1,... \} $. Each of the $N = 2m + 2$ transitions that occur happens at a particular time $t_i$, where $t_{2m+2} = T$ (see Figure \ref{fig:schematic}(b)). The probability density of this path may be written as:

\begin{multline}
P(m,T,\{t_i\}) = P_{d}(t_1)\left[ P_{r}(t_2-t_1)S_{s}(t_2-t_1) \right]... \\ \times \left[ P_{s}(T - t_{N-1})S_{r}(T - t_{N-1}) \right],
\end{multline}

\noindent where $S_{x}(t) = 1 - \int_0^{t} dt^{\prime}\mathop{} P_{x}(t^{\prime})$ is the \textit{survival probability} associated with the waiting time distribution $P_{x}$ \cite{VanKampen}. Marginalising over the $t_i$, $1 \leq i < N$, while enforcing the ordering of these intermediate times (as, in a similar context, in \cite{SanjibDeviations}), we obtain the joint distribution of $T$ and $m$:

\be
P(m, T)=\int_0^T dt_{N-1} \mathop{} \ldots\int_0^{t_2} dt_1 \mathop{} P(m,T,\{t_i\}).
\ee

\indent The Laplace transform of $P(m,T)$ can be compactly written as follows:

\be \label{eq:starting}
\mathcal{L}\l P(m,T) \r \equiv \int_0^{\infty} dT \, P(m,T)\,\me^{-st} = \hat{G}_f(s) \l \hat{G}_i(s)\r^m
\ee

\noindent where $\hat{G}_i(s) = \Lagr \left[ P_{d} (t) \right] \times  \Lagr \left[ P_{r} (t) S_{s} (t) \right]$ and $\hat{G}_f(s) = \Lagr \left[ P_{d} (t) \right] \times \Lagr \left[ P_{s} (t) S_{r} (t) \right]$. Note that setting $s$ to $0$ in Eq. (\ref{eq:starting}) corresponds to marginalising $P(m,T)$ over $T$, yielding $P(m)$. Similarly, we obtain $P(T)$ by marginalising over $m$:

\be \label{eq:disorder_free}
\mathcal{L}\l P(T) \r = \hat{G}_f \sum_{m=0}^{\infty} \hat{G}_i^m = \f{\hat{G}_f(s)}{1-\hat{G}_i(s)}.
\ee

\indent Eqs. (\ref{eq:starting}) and (\ref{eq:disorder_free}) constitute the foundational results of our paper. 
Each $G_x^{(n)}$, and therefore $P(T)$, can be computed from the underlying waiting time distributions, either analytically or by quadrature.
Moreover, the moments of the distribution can be obtained directly from derivatives of $\Lagr \l P(T) \r$ at $s = 0$. When the transition from $A \to B$ is instananeous (the calculation without this assumption is similar, albeit more tedious, as is the calculation of higher moments), we may expand $\hat{G}_f$ and $\hat{G}_i$ as $\hat{G}_x(s) = G^{(0)}_x - G^{(1)}_x \, s + \f{1}{2} G^{(2)}_x \, s^2$, where

\be \label{eq:expandedprop}
G^{(n)}_x = \int dt \, t^n \, G_x(t),
\ee

\noindent and $G_i(t) = P_{r} (t)~S_{s} (t)$ and $G_f(t) = P_{s} (t)~S_{r} (t)$. The normalisation $\int P(T)dT = 1$ implies that $G_f^{(0)} = 1 - G_i^{(0)}$. Note that $G^{(0)}_i$ and $G^{(0)}_f$ admit a simple interpretation as the splitting probabilities for restarting (transition $B \to A$) and search completion (transition $B \to C$), respectively, from state $B$ in Figure \ref{fig:schematic}. $G^{(n)}_x/G^{(0)}_x$ is then the $n$th moment of the respective conditional exit time distributions \cite{VanKampen}.

\indent Inserting (\ref{eq:expandedprop}) into (\ref{eq:disorder_free}) and expanding to first and second order in $s$ we find:

\bea \label{eq:FPTmeanAndVar}
&& \langle T \rangle = \f{G^{(1)}_f + G^{(1)}_i}{G^{(0)}_f} \nn, \\
&& \langle T^2 \rangle = \f{2G^{(1)}_i \l G^{(1)}_f + G^{(1)}_i \r + G^{(0)}_f \l G^{(2)}_f + G^{(2)}_i \r}{(G^{(0)}_f)^2}.
\eea

\begin{figure*}
\begin{centering}
\includegraphics[width=17cm]{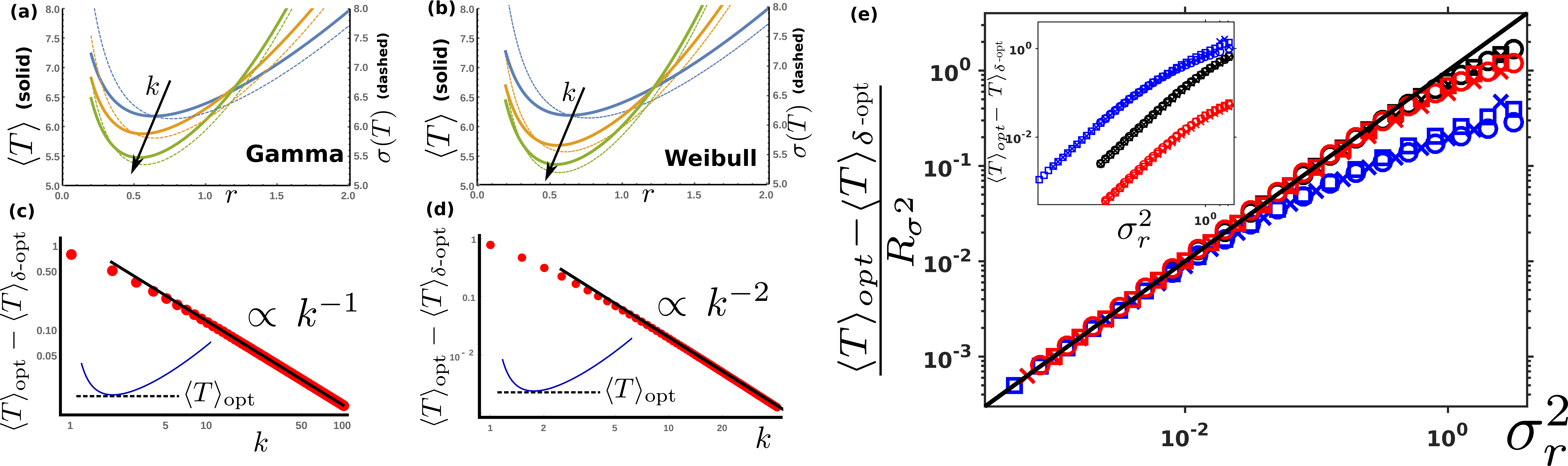}
\caption{\label{fig:whirly} The mean (solid line) and standard deviation (dashed line) of the completion time for Gamma \textbf{(a)} and Weibull \textbf{(b)} restarting with mean restart time $1/r$, with shape parameters $k = 1$ (blue), $k = 2$ (orange) and $k = 10$ (green). \textbf{(c)} and \textbf{(d)} show an algebraic decay of the optimal completion time ($\langle T \rangle_{\text{opt}}(k)$), for Gamma and Weibull restarting respectively, as a function of shape parameter $k$. \textbf{(e)} Optimal mean completion time $\langle T \rangle_{\text{opt}}$, scaled by the response factor $R_{\sigma^2}$ (see Eq. (\ref{eq:Topt_scaling_factor})), as a function of the variance of the restart distribution. Solid line is a straight line with slope $1$ and intercept $0$. Inset shows the same without the scaling. Search distributions $P_s(t)$ shown are: Levi-Smirnov (black), log-normal (red) and Frechet (blue); for parameter values see supplementary information \cite{SI}. Restart distributions $P_r(t)$ are: Gamma (crosses), Weibull (circles) and truncated normal (squares). Data points were generated by quadrature and numerical optimisation using Eqs. (\ref{eq:expandedprop}) and (\ref{eq:FPTmeanAndVar}).}
\end{centering}
\end{figure*}

\indent Eqs. (\ref{eq:disorder_free}) and (\ref{eq:FPTmeanAndVar}) generalise the expressions found in \cite{ReuveniUniversal} to arbitrarily distributed restarts. We mentioned earlier that the mean completion time may be finite even if, without restarts, it diverges. Remarkably, this may be true even when the mean of the \textit{restart} distribution also diverges. One such instance is a 1D diffusive search with disordered, random restarting; that is, restarts that occur at a constant rate $k$ that is itself drawn (after every restart event) from, for example, an exponential distribution with mean $1/\eta$. The disorder-averaged restart distribution then decays algebraically $\propto t^{-2}$, such that the average time between restarts diverges. Nonetheless, each $G^{(1)}_{i/f}$ is finite and, from Eq. (\ref{eq:FPTmeanAndVar}), so is the mean completion time \cite{SI}.

\indent A particularly interesting limit of Eq. (\ref{eq:FPTmeanAndVar}) is one in which the restart occurs deterministically at a time $\tau$, for which the mean completion time may be computed to be (as found by other means in \cite{Redner,PalTimeDepResets,PalReuveni}):

\be \label{eq:deltareset}
\langle T \rangle_{\delta} =  \f{\int_0^{\tau} dt \, S_s(t)}{1- S_s(\tau)}
\ee

\indent This expression is finite for any $\tau$ for which $S_s(\tau) < 1$, which leads to the remarkable conclusion that a deterministic restart (optimal or not) may \textit{always} render the mean completion time finite. Further, $\langle T \rangle_{\delta\text{-opt}} \equiv \min_\tau \langle T \rangle_{\delta}$ is in fact the \textit{globally} optimal restart time, over all restart distributions \cite{PalTimeDepResets,PalReuveni}, which is achieved at the restart time $\tau_{\text{opt}} = \argmin_\tau \langle T \rangle_{\delta}$. How close to this optimum does a stochastic restart mechanism get? To understand the landscape of optimal completion times, we go on to ask how the introduction of fluctuations to a deterministic restart mechanism (that is, varying the shape of the restart distribution `away' from a $\delta$-function) affects the optimality of the completion time.

\indent To do so in a consistent manner, we recall that several families of distributions interpolate smoothly between a $\delta$-function (deterministic restarting) and an exponential distribution (Poisson restarting). Two illustrative examples are the Gamma family of distributions, $\f{(k r)^k t^{k-1}}{\Gamma(k)} \me^{-krt}$, and the Weibull family, $\f{k}{\lambda^k} t^{k-1}\exp \l -\l \f{t}{\lambda} \r ^k \r $ with $\lambda = 1/ r \, \Gamma \l 1 + 1/k \r$, such that both distributions have mean $1/r$. Each of these families is parameterised by a `shape factor' $k$ such that they are exponential for $k=1$ and approach a $\delta$-function as $k \to \infty$.

\indent We illustrate the effect of the shape factor $k$ on the mean completion time of a diffusive search in Figure \ref{fig:whirly}(a-b). We observe that the optimal value of $\langle T \rangle$ decreases with $k$ \cite{ShapeMinimum}, approaching $\langle T \rangle_{\delta\text{-opt}}$ as a power law $k^{\alpha}$, with numerically determined exponents $\alpha \approx -1$ for Gamma restarting and $\alpha \approx -2$ for Weibull restarting (Figure \ref{fig:whirly}(c-d)).

\indent These algebraic decays, while at first appearing to be of mysterious origin, in fact also describe the variance of a Gamma or Weibull distribution as $k \to \infty$ and $r$ is held fixed \cite{SI}. This suggests that as the restart grows more `punctual' (i.e. the restart distribution becomes more tightly distributed, approaching a $\delta$-function), the optimal completion time approaches $\langle T \rangle_{\delta\text{-opt}} \propto \sigma_r^2$, the variance of $P_r(t)$. We show that this is true for the general restart problem by exploiting a central moment representation to expand a generic $P_r(t)$ around its mean $\tau$, valid for any distribution with finite moments \cite{GillespieExpansion}:

\be \label{eq:gill_expansion}
P_r(t) = \delta \l t - \tau \r + \sum_{n = 2}^{\infty} \f{(-1)^n}{n!} \mu_n(\tau) \, \delta^{(n)} \l t - \tau \r
\ee

\noindent where $\mu_n(\tau)$ is the $n$th central moment of the restart distribution: $\int dt \, (t-\tau)^n P_r(t)$. In particular, $\mu_2(\tau) = \sigma_r^2$, the variance of the restart distribution. As the distribution grows more peaked, we may neglect higher moments and truncate the expansion to the term $\propto \sigma_r^2$. Inserting the expansion into Eq. (\ref{eq:FPTmeanAndVar}) we find $\langle T\rangle = \langle T \rangle_{\delta}(\tau) + R_{\sigma^2}(\tau)\sigma_r^2+ \mathcal{O}(\mu_3 (\tau))$, where $R_{\sigma^2}$, which determines the response of the completion time to fluctuations of the restart process, is:

\be \label{eq:Topt_scaling_factor}
R_{\sigma^2} = -\frac{\langle T \rangle_{\delta}}{2\left(1-S_s(\tau)\right)} \, \left[\frac{P_s(\tau)}{\langle T \rangle_{\delta}}+{\f{\p P_s}{\p t}\vline}_{\,t=\tau}\right]
\ee

\indent This expression, which depends \textit{only} on the search distribution $P_s(t)$, is valid for any punctual restart distribution -- in particular, we have not yet demanded that the mean restart time $\tau$ be optimal. If we now suppose that the optimal mean restart time is approximately constant for small $\sigma_r^2$, then the \textit{optimal} $\langle T \rangle$ will increase linearly with $\sigma_r^2$ with slope $R_{\sigma^2}$. This is borne out by Figure \ref{fig:whirly}(e), in which we observe that, as $\sigma_r^2 \to 0$, $\langle T \rangle_{\text{opt}}$ approaches the respective global optimum $\langle T \rangle_{\delta\text{-opt}}$ as predicted by Eq. (\ref{eq:Topt_scaling_factor}).

\indent Inspection of Eq. (\ref{eq:Topt_scaling_factor}) further reveals that fluctuations in a (possible non-optimal) restart process do not always increase the mean completion time \cite{EuleMetzger,ShapeMinimum}. However, as we know that optimal deterministic restarts ($\tau = \tau_{\text{opt}}$) are indeed optimal in the entire space of restart distributions, it must be true that $R_{\sigma^2} > 0$ for $\tau = \tau_{\text{opt}}$. This implies that $\p_t P_s(\tau_{\text{opt}}) < 0$, which provides for a rather simple lower bound on $\tau_{\text{opt}}$ for singly peaked search distributions $P_s(t)$ (such as those considered here): $\tau_{\text{opt}}$ must lie to the right of the mode of the distribution; e.g., $\tau_{\text{opt}}>\alpha^2/3$ for Levi-Smirnov (where $\alpha$ is the initial distance to target), $\tau_{\text{opt}}>(\alpha/(1+\alpha))^{1/\alpha}$ for a one parameter Frechet distribution, $\tau_{\text{opt}}>\alpha((\beta-1)/(\beta+1))^{1/k}$ for log-logistic, and $\tau_{\text{opt}}>e^{-\sigma^2}$ for a log-normal distribution \cite{DistributionDefinition}. 

\indent Up until now, we have considered an `optimal' protocol to be one that minimises the mean completion time of the search. However, the efficacy of a search protocol may not be determined by the mean completion time but instead by other constraints; for instance, constraints on the number of restart events before the search concludes. These may be in terms of a time overhead \cite{ReuveniUniversal}, which would contribute to the delay distribution $P_d(t)$, or an energetic or financial cost. We suppose that each restart event incurs a fixed cost $\gamma$, and thus consider a \textit{cost function} linear in $m$: $f(m) = \gamma\, m$. Marginalising Eq. (\ref{eq:starting}) over $T$ and then averaging: $\langle f(m) \rangle \propto \langle m \rangle = {G^{(0)}_i}/{G^{(0)}_f}$.

\begin{figure}
\begin{centering}
\includegraphics[width=8.5cm]{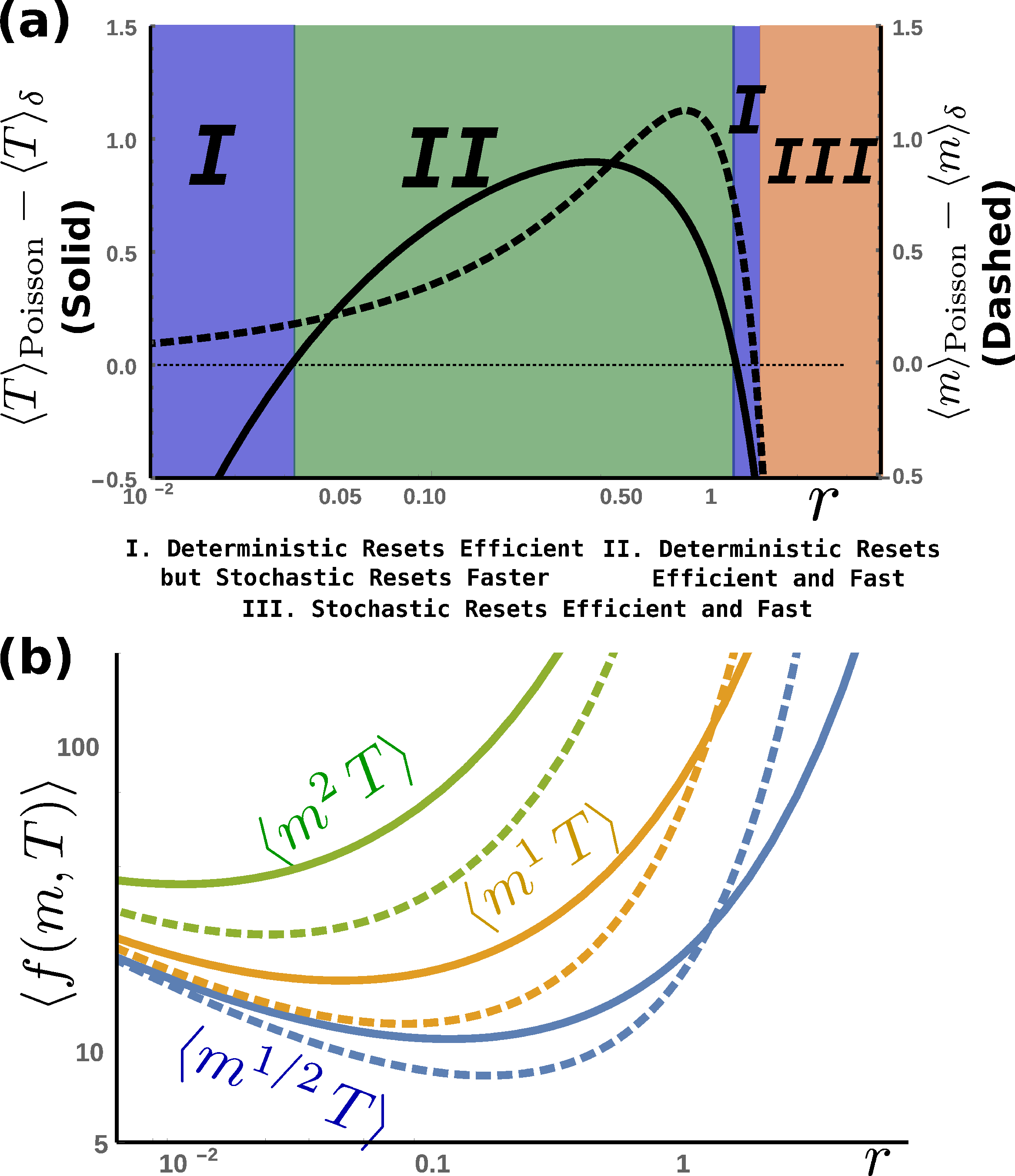}
\caption{\label{fig:costs} \textbf{(a)} The average number of restart events for 1D diffusion with Poisson (solid) and deterministic (dashed) restarts, as a function of the inverse average restarting time (`rate') $r$. \textbf{(b)} The evaluated cost functions $f(m,T) = \sqrt{m}\,T$ (blue), $m\, T$ (orange) and $m^2\,T$ (green) for 1D diffusion with Poisson (solid) and deterministic (dashed) restarting with mean $1/r$.}
\end{centering}
\end{figure}

\indent We may now study the cost incurred by different restart protocols. Evaluating $\langle m \rangle$ for 1D diffusion with Poisson or deterministic restarts, we find that deterministic restarts incur a lower cost for restart rates $r$ less than $\approx 1.412$, but at higher $r$ the cost rapidly outpaces that of the stochastic restart mechanism \cite{SI}. In conjunction with Eqs. (\ref{eq:FPTmeanAndVar}) and (\ref{eq:deltareset}), we may thereby identify regimes corresponding to a tradeoff between efficiency (minimising $\langle m \rangle$) and speed (minimising $\langle T \rangle$) that depend on the nature of the restart protocol (Figure \ref{fig:costs}(a)).

\indent Finally, we consider a case in which one might be interested in \textit{simultaneously} minimising the number of restarts and the completion time. We must construct an appropriate cost function, $f(m,T)$ that depends on both $m$ and completion time $T$, and then average it over the joint distribution $P(m,T)$, Eq. (\ref{eq:starting}). We consider the form $f(m,T) = m^{\beta} T$, with a relative weight between time and efficiency given by the exponent $\beta$. Using Eqs. (\ref{eq:starting}) and (\ref{eq:expandedprop}), we find for $\langle m^{\beta}T \rangle$:

\be
\frac{1}{\langle m \rangle} G^{(0)}_f G_i^{(1)} \Phi \l G_i^{(0)},-\beta -1,0 \r +G_f^{(1)} \Phi \l G_i^{(0)},-\beta ,0 \r
\ee

\noindent where $\Phi(a,b,c)$ is the Hurwitz-Lerch transcendent. This is plotted for various values of $\beta$ in Figure \ref{fig:costs}(b), once again for the case of 1D diffusion with Poisson or deterministic restarts, showing that deterministic restarting continues to be the more `optimal' protocol.

\indent To summarize, we have introduced a simple and general framework to analyse stochastic searches with restarts, that works for arbitrary search and restart processes (and indeed, arbitrary delays $P_d(t)$). This allows us to analytically calculate (or reduce to quadratures) the moments of completion time, number of restarts, and a variety of cost functions combining these two. Thereby we are able to derive several results on the optimality of a wide range of search and restart processes. In particular, we have shown that (i) the peak of the search distribution $P_s(t)$ provides a simple lower bound for the optimal deterministic restart time, $\tau_{\text{opt}}$, (ii) the mean completion time of a search scales linearly with the variance of a punctual restart distribution, and (iii) deterministic restarting may invoke higher operating costs than stochastic mechanisms. Our calculations, complementary to \cite{MMRS_Reuveni,ReuveniUniversal,PalReuveni}, provide powerful tools with which to characterise the completion times of a large class of systems with restarts, and will aid the rational design of optimal, efficient restart mechanisms.

\indent \textbf{Acknowledgements} We are grateful to Amit Kumar Singh, Mogens Jensen, Richard Morris and Martin Evans for their helpful comments, and B.R. Ujwal and T.E. Amit for technical advice. This work was supported by the Simons Foundation.


\begin{thebibliography}{30}
\bibitem{ErlangWeibull} The Gamma distribution with integer shape factor $k > 0$ and mean $1/r$ may be derived as the waiting time for $k$ sequential Poisson events, each with rate $r/k$. Non-integral $k$ follows by analytic continuation. The Weibull distribution may be similarly derived as the waiting time distribution for a time-varying rate $\propto t^{k-1}$.
\bibitem{ShapeMinimum} Note that $\langle T \rangle$ may behave non-monotonically with $k$ when $r$ is held fixed, as found by simulation in \cite{EuleMetzger}.
\end{thebibliography}
\end{document}


\maketitle

\section{1D Diffusive Search}

\indent The first passage time distribution $P(t)$ for a 1D diffusive search is given by the Levi-Smirnov distribution:

\be \label{eq:levismirnov}
P(t) = \f{x_0}{\sqrt{4 \pi D \, t^3}} \me^{-\f{x_0^2}{4Dt}}
\ee

\noindent where $x_0$ is the distance to the target and $D$ is the diffusion coefficient. We choose as the unit of time $x_0^2/4D$, with which the distribution takes the appealingly simple form $\f{1}{\sqrt{\pi}\,t^{3/2}} \, \me^{-1/t}$. Note that this is a different choice than the one made by Evans and Majumdar in \cite{EvansMajumdarPRL}.

\section{Diffusion with `Disordered' Restarts}

\indent In the main text we construct an example in which the mean of both the search and restart distributions diverge but the mean completion time remains finite: a 1D diffusive search with Poisson restarts whose rate is drawn from an exponential distribution with mean $1/\eta$. As, to the best of our knowledge, this calculation has not been presented before, we do so here.

\indent The restart rate $k$ is drawn anew after each restart from the disorder distribution $P(k) = \eta \, \me^{-\eta \, k}$ (corresponding to a kind of `annealed' disorder). The characteristic time-scale of restarting is then $\eta$, with $1/\eta$ being the characteristic `rate'. Denoting an average over $P(k)$ by an overbar, we find $\overbar{P}_r(t) = \eta/\l \eta + t \r^2$ and $\overbar{S}_r(t) = \eta/\l \eta + t \r$. Inserting these into the definitions of $G_i(t)$ and $G_f(t)$, we find for the mean completion time:

\be
\langle T \rangle = \frac{\pi  \eta \,  \text{erfi}\left(\frac{1}{\sqrt{\eta }}\right)-2 \, _2F_2\left(1,1;\frac{3}{2},2;\frac{1}{\eta }\right)}{1-\frac{\sqrt{\pi } e^{\frac{1}{\eta }} \text{erfc}\left(\frac{1}{\sqrt{\eta
   }}\right)}{\sqrt{\eta }}}
\ee

\begin{figure*}
\begin{centering}
\includegraphics[width=15cm]{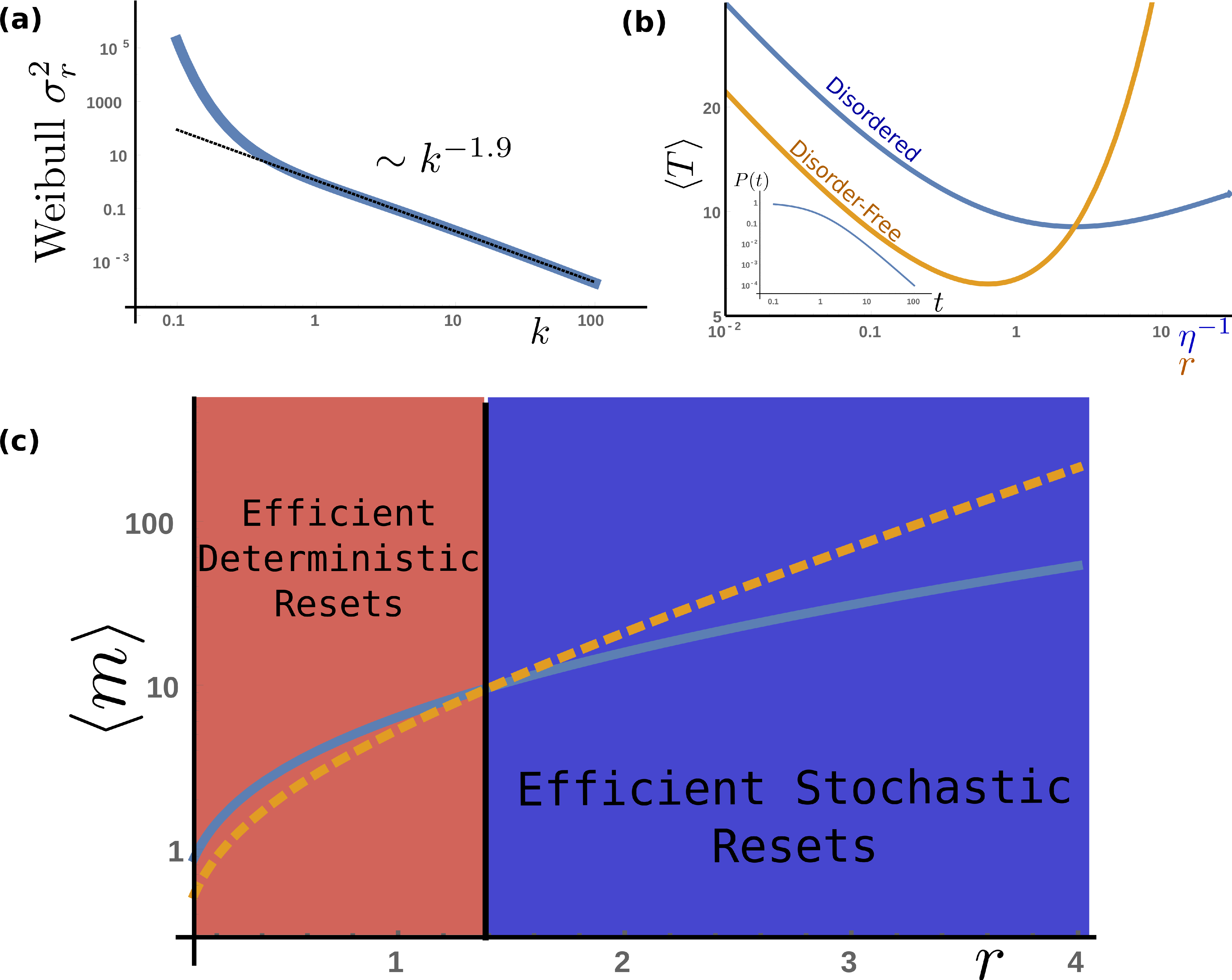}
\caption{\label{fig:si1} \textbf{(a)} Variance of a Weibull distribution $\sigma_r^2$ as as function of the shape factor $k$. Dotted line is $\sim k^{-1.9}$. \textbf{(b)} Average completion time $\langle T \rangle$ for Poisson (orange) restarts with rate $k$ and exponentially disordered Poisson restarts (blue) with characteristic restart time scale $\eta$. Inset shows disorder averaged restart time distribution $\overbar{P}_r(t)$ - i.e., the actual waiting time between restart events. Note the algebraic decay $\propto t^{-2}$  \textbf{(c)} Plot of $\langle m \rangle$ for a 1D diffusive search with Poisson (solid) and deterministic (dashed) restarts, as a function of the inverse mean restart time, $r$.}
\end{centering}
\end{figure*}

\indent We plot this, as a function of $\eta^{-1}$, against the disorder-free case (simple Poisson restarts) in Figure \ref{fig:si1}(b).

\section{Gamma and Weibull Distributions}

\indent In the text we analyse the properties of a diffusive search subject to restarts according to a Gamma ($P_r(t) = \f{(k r)^k t^{k-1}}{\Gamma(k)} \me^{-krt}$) or Weibull ($\f{k}{\lambda^k} t^{k-1}\exp \l -\l \f{t}{\lambda} \r ^k \r$) distribution. Note that the mean of the Gamma distribution is $1/r$ and hence varying $k$ while keeping $r$ fixed is easily accomplished. To facilitate comparison between the distributions, we reparameterise the Weibull distribution by the substitution $\lambda = \f{1}{r \, \Gamma \l 1+ 1/k \r}$, where the mean of the Weibull distribution is now also $1/r$.

\indent The mean and standard deviations for each $k$ were then calculated from the definitions of the $G^{(n)}_x$ given in the main text, either analytically (for the Gamma distribution) or by quadrature (for the Weibull distribution).

\indent In the main text we remark that the variance of these distributions depend as power laws on the shape factor $k$ as $k \to \infty$. We demonstrate this here. The variance of the Gamma distribution is simply $\sigma_r^2 = 1/k\,r^2$, from which the claimed dependence can be immediately seen. For the Weibull distribution, the variance is:

\be \label{eq:weibullvariance}
\sigma_r^2 =  \l \f{1}{r \, \Gamma \l 1+ 1/k \r} \r^2 \left[ \Gamma \l 1+ \f{2}{k} \r - \l  \Gamma \l 1 + \f{1}{k} \r  \r^2 \right]
\ee 

\indent This is plotted in Figure \ref{fig:si1}(a) for a fixed $r$. We see that for large $k$ the variance behaves algebraically with exponent $\approx - 1.9$. This was identical to the numerically determined exponent for $\langle T \rangle_{\text{opt}} - \langle T \rangle_{\delta\text{-opt}}$, which we reported in the main text as $\approx -2$ for simplicity.

\section{Numerical Procedure for $\langle T \rangle_{\text{opt}} - \langle T\rangle_{\delta\text{-opt}}$ vs. $\sigma_r^2$ Plot}

For each pair of search and restart distributions, we used quadrature to find the value of $\langle T \rangle_{\text{opt}}$. This was done by exploting the two-parameter nature of the restart distributions chosen. The distributions were first reparameterised by the mean $1/r$ and the variance $\sigma_r^2$. Then, for each value of $\sigma_r^2$, quadrature was used to find the mean completion time $\langle T \rangle$ for a given $r$ -- this was then numerically optimised over $r$ to find $\langle T \rangle_{\text{opt}}$. This was repeated for each value of $\sigma_r^2$, values of which were chosen so as to be uniformly distributed in log-space.

\indent The search distributions chosen for this calculation were: Levi-Smirnov: $\f{1}{\sqrt{\pi}\,t^{3/2}} \, \me^{-1/t}$, Frechet: $\f{1}{2\, t^{3/2}} \exp \l -1/\sqrt{t} \r$ and log-normal: $\f{1}{\sqrt{2 \pi}\, t} \exp \l -(\ln t)^2/2 \r$

